\documentclass[a4paper]{article}
\usepackage{graphicx}
\usepackage{mathrsfs} 
\usepackage{amsmath,amssymb}

\def\bvec#1{\mbox{\boldmath $#1$}}

\begin{document}


\title{\bf Zeeman effects on the entanglement of non-equilibrium finite-spin systems }


\author{Koichi Nakagawa \\[3mm] \small Laboratory of Physics,  \small School of Pharmacy and Pharmaceutical Sciences, \\ \small Hoshi University, Tokyo 142-8501, Japan}


\date{\empty}


\maketitle

\begin{abstract}
We study the Zeeman effect on entanglement of non-equilibrium finite-spin systems with external fields using a method based on thermofield dynamics (TFD). For this purpose, the extended density matrices and extended entanglement entropies of two systems with either non-competing or competing external fields are calculated according to the dissipative von Neumann equation, and the numerical results are compared. Consequently, through the ``twin-peaks'' oscillations of the quantum entanglement, we have illustrated the Zeeman effect on the entanglement of non-equilibrium finite-spin systems with competing external fields in the TFD algorithm. 
\end{abstract}
\vspace{5mm}
\section{Introduction \label{sec:1}}
The entanglement of quantum states is a correlation between multiple systems that is peculiar to quantum mechanics \cite{Peres}. The behavior of entangled quantum states is called ``quantum entanglement''. It plays important roles in quantum computation and quantum information \cite{Nielsen}, and is useful in applications of the AdS/CFT correspondence \cite{Ryu,Nishioka}. Entanglement entropy is a measure of the strength of quantum entanglement. It has been used as an order parameter of quantum spin systems \cite{Stephan,Tanaka}.

A different method to analyze quantum entanglement using thermofield dynamics (TFD) \cite{Fano,Prigogine,Takahashi} was proposed in Ref.~\cite{Hashizume}, and its applicability confirmed in Ref.~\cite{Nakagawa}. In this treatment of quantum entanglement with TFD, an extended density matrix is defined on the doubled Hilbert space (physical and ancillary Hilbert spaces), and examined for some simple cases~\cite{Hashizume}. The TFD-based method allows the entanglement states to be easily understood, because the intrinsic elements caused by quantum entanglement can be extracted from the extended density matrix in this formulation. Consequently, it was found that the intrinsic quantum entanglement can be distinguished from the thermal fluctuations included in the definition of ordinary quantum entanglement at finite temperatures. Based on the analysis presented in Ref.~\cite{Hashizume}, it was argued that the general TFD formulation of the extended density matrix is applicable, not only to equilibrium states, but also to non-equilibrium states. The extended density matrix was calculated as a simple example in Ref.~\cite{Hashizume} for the case of equilibrium finite-spin systems with external fields. In Ref.~\cite{Nakagawa}, it was shown that the extended entanglement entropy, which was obtained by using the extended density matrix, could be decomposed into parts from thermal (but classical) fluctuations and from quantum entanglement, as in Eqs.\, \eqref{eq:23}$\sim$\eqref{eq:23b}. In the present communication, we prove that the value of the extended entanglement entropies of the non-equilibrium finite-spin systems is positive semi-definite. We then obtain the extended entanglement entropies of TFD as a well-defined measure of quantum entanglement.

Moreover, in Ref.~\cite{Hashizume}, the authors gave an example of a frustration effect on the entanglement of equilibrium finite-spin systems that is caused by competing external fields. They compared a system that does not contain a competing effect between the interaction and the external fields with one that does contain it. They concluded that this competing effect is a kind of frustration, in relation to the partial recovery of the broken symmetry of the spin inversion, by using the equilibrium density matrix. However, in Ref. \cite{Hashizume}, because only a two-spin system is considered, we feel that it is incorrect to refer the effect of external fields as ``frustration''. In addition, the effect of the external fields was analyzed only in an equilibrium two-spin system, and not in a non-equilibrium two-spin system \cite{Hashizume}. The effects of external magnetic fields have generally been considered among the Zeeman effects.
Thus, the Zeeman effect on the entanglement of non-equilibrium systems with external fields remains of interest. In the present communication, we therefore investigate, exhaustively, the extended density matrices and entanglement entropies of non-equilibrium spin systems with both non-competing and competing external fields based on the general TFD algorithm. By comparing the quantities for the non-competing and competing cases, we demonstrate the Zeeman effect on the dissipative dynamics of the entanglement through ``twin-peaks'' oscillations of the extended density matrices and the extended entanglement entropies. Furthermore, the origin and the generality of the ``twin-peaks'' oscillations are argued by using the extended entanglement entropies. 

This paper is organized as follows: we introduce the extended density matrices of the non-equilibrium systems with external fields in the next section and examine their properties. In section 3, we obtain the extended entanglement entropies of non-equilibrium spin systems with competing and non-competing external fields, and discuss the numerical results. The last section is devoted to discussion and conclusions.

\section{Extended density matrices of non-equilibrium finite-spin systems with external fields \label{sec:2}}

The Zeeman Effects was originally studied in the Ising spin-glass model, which is described by the Hamiltonian
\begin{align}
\mathcal{H}=-\sum_{\langle i,\,j\rangle}J_{ij}\bvec{S}_{i}\cdot \bvec{S}_{j}-\sum_{i}\bvec{h}_{i}\cdot \bvec{S}_{i},
\label{eq:last}
\end{align}
where $J_{ij}$ is the strength of spin interaction, $\bvec{h}_{i}$ is the external magnetic field, and $\langle i,\,j\rangle$ express the pairs of nearest neighbors. Here, the Hamiltonian is the most simple separation of Eq.\,\eqref{eq:last}. 
Consider the $S=1/2$ spin system described by the Hamiltonian~\cite{Hashizume}
\begin{align}
\mathcal{H}:=-J\bvec{S}_{\text{A}}\cdot \bvec{S}_{\text{B}}-g\mu_B\left(H_AS_{\text{A}}^z+H_BS_{\text{B}}^z\right) , 
\label{eq:1}
\end{align}
which involves the spin operators, $\bvec{S}_{\text{A}}=(S_{\text{A}}^x, S_{\text{A}}^y, S_{\text{A}}^z)$ and $\bvec{S}_{\text{B}}=(S_{\text{B}}^x, S_{\text{B}}^y, S_{\text{B}}^z)$ of the subsystems A and B, respectively, where $g$ is the Lande factor, $\mu_B$ the Bohr magneton, and $H_{\rm A}$ and $H_{\rm B}$ are the components of the external fields conjugate to $\bvec{S}_{\text{A}}$ and $\bvec{S}_{\text{B}}$, respectively. To examine the Zeeman effect on the entanglement, Ref.~\cite{Hashizume} recognized that the two systems with Hamiltonians, $\mathcal{H}^{\rm nc}$ and $\mathcal{H}^{\rm c}$, should be compared for the cases of a non-competing external field $H_{\rm A}=H_{\rm B}=H$ and competing external field $H_{\rm A}=-H_{\rm B}=H$, respectively. As can also be seen from Eq.~\eqref{eq:1}, the spin inversion symmetry of these systems is broken by these external fields in the Hamiltonian. In a forthcoming analysis exploring the Zeeman effect on the dissipative dynamics of entanglement, the partial recovery of this broken symmetry will play an important role.

The state, $|s\rangle$, of the total system is defined by the direct product, $|s\rangle=|s_{\text{A}},s_{\text{B}}\rangle=|s_{\text{A}}\rangle |s_{\text{B}}\rangle$. Using the base $\left\{ |++\rangle, |+-\rangle, |-+\rangle, |--\rangle \right\} $, the matrix form of the Hamiltonian \eqref{eq:1} is expressed as 
\begin{align}
\mathcal{H}^{\rm nc}&=-J\bvec{S}_{\text{A}}\cdot \bvec{S}_{\text{B}}-g\mu_BH\left(S_{\text{A}}^z+S_{\text{B}}^z\right) \nonumber \\ 
&=\left( -\frac{J}{4}-g\mu_BH \right)|++\rangle\langle++|+\left( -\frac{J}{4}+g\mu_BH \right)|--\rangle\langle--| \nonumber \\
&\hspace{3mm}+\frac{J}{4}\left( |+-\rangle\langle+-| +|-+\rangle\langle-+|\right)-\frac{J}{2}\left(
|+-\rangle\langle-+| +|-+\rangle\langle+-|\right)
\label{eq:1nc}
\end{align}
for the non-competing case and
\begin{align}
\mathcal{H}^{\rm c}&=-J\bvec{S}_{\text{A}}\cdot \bvec{S}_{\text{B}}-g\mu_BH\left(S_{\text{A}}^z-S_{\text{B}}^z\right)\nonumber \\ 
&=-\frac{J}{4}\left( |++\rangle\langle++|+|--\rangle\langle--| \right)  \nonumber \\
&\hspace{3mm}+\left( \frac{J}{4}-g\mu _BH \right)  |+-\rangle\langle+-| +\left( \frac{J}{4}+g\mu _BH \right) |-+\rangle\langle-+| \nonumber \\ &\hspace{3mm}-\frac{J}{2}\left(
|+-\rangle\langle-+| +|-+\rangle\langle+-|\right) \label{eq:1a}
\end{align} 
for the competing case.

Next, consider non-equilibrium systems with dissipation, described by the Hamiltonian of Eq.~\eqref{eq:1}. The time dependence of the ordinary density matrix, $\rho ^{\alpha }(t)$, of these systems is given by the dissipative von Neumann equation \cite{Suzukibook,Suzuki1} 
\begin{equation}
i\hbar\frac{\partial}{\partial t}\rho ^{\alpha }(t)=[\mathcal{H}^{\alpha },\rho ^{\alpha }(t)]-\epsilon \left( \rho ^{\alpha }(t)-\rho ^{\alpha }_{\text{eq}}
\right),\label{eq:2}
\end{equation}
where $\epsilon $ is a dissipation parameter, $\alpha $ ``nc'' for the non-competing case or ``c'' for the competing case, and $\rho ^{\alpha }_{\text{eq}}$ the ordinary density matrix of the equilibrium systems \cite{Hashizume}.
The solution of Eq.~\eqref{eq:2} is then expressed as
\begin{equation}
\rho ^{\alpha }(t)=e^{-\epsilon t}{U^{\alpha }}^{\dagger}(t)\rho_0 U^{\alpha }(t)+(1-e^{-\epsilon t})\rho ^{\alpha }_{\text{eq}},\label{eq:3}
\end{equation}
for an arbitrary initial density matrix, $\rho_0$, where the unitary operator, $U^{\alpha }(t):=e^{i\mathcal{H}^{\alpha }t/\hbar}$, denotes
\begin{align}
U^{\rm nc}(t) &=e^{i\omega t/4} \left( \exp{\left(\frac{-i(\omega +2g\mu_BH/\hbar)t}{2}\right)}|++\rangle\langle++| \right.\nonumber \\
&\hspace{13.5mm}+\exp{\left(\frac{-i(\omega -2g\mu_BH/\hbar)t}{2}\right)}|--\rangle\langle--| \nonumber \\ 
&\hspace{13.5mm}+\cos \frac{\omega t }{2}\left( |+-\rangle\langle+-|+|-+\rangle\langle-+| \right) \nonumber \\ 
&\hspace{13.5mm}-\left.i\sin \frac{\omega t }{2}\left( |-+\rangle\langle+-|+|+-\rangle\langle-+| \right) \right) \label{eq:7nc}
\end{align}
for the non-competing case and
\begin{align}
U^{\rm c}(t) &=e^{i\omega t/4} \left( e^{- i \omega t /2}\left( |++\rangle\langle++| +|--\rangle\langle--| \right) \rule{0mm}{10mm}\right. \nonumber \\ 
&+ \left( \cos \left(\frac{t \sqrt{4 g^2H^2 \mu _B^2+\omega ^2 \hbar ^2}}{2 \hbar }\right)-\frac{2 i gH \mu _B \sin \left(\frac{t \sqrt{4 g^2H^2 \mu _B^2+\omega ^2 \hbar ^2}}{2 \hbar }\right)}{\sqrt{4 g^2H^2 \mu _B^2+\omega ^2 \hbar ^2}} \right) \nonumber \\ 
&\times |+-\rangle\langle+-|\nonumber \\ 
&+\left( \cos \left(\frac{t \sqrt{4 g^2H^2 \mu _B^2+\omega ^2 \hbar ^2}}{2 \hbar }\right)+\frac{2 i gH \mu _B \sin \left(\frac{t \sqrt{4 g^2H^2 \mu _B^2+\omega ^2 \hbar ^2}}{2 \hbar }\right)}{\sqrt{4 g^2H^2 \mu _B^2+\omega ^2 \hbar ^2}} \right) \nonumber \\ 
&\times |-+\rangle\langle-+| \nonumber \\ 
&\left.-\frac{i \omega  \hbar  \sin \left(\frac{t \sqrt{4 g^2H^2 \mu _B^2+\omega ^2 \hbar ^2}}{2 \hbar }\right)}{\sqrt{4 g^2H^2 \mu _B^2+\omega ^2 \hbar ^2}}\left( |-+\rangle\langle+-|+|+-\rangle\langle-+| \right) \right) 
\label{eq:7}
\end{align}
for the competing case. Here, $\omega:= J/\hbar$. The explicit expression for $\rho ^{\alpha }(t)$ in Eq.~\eqref{eq:3} is complicated for an arbitrary initial condition and it is difficult to understand its physical meaning. Hence, we will, for the sake of simplicity, hereafter confine the discussion to the initial condition $\rho_0=|+-\rangle\langle+-|$. Inserting Eq.~\eqref{eq:7nc} or \eqref{eq:7}, along with the initial condition, into Eq.~\eqref{eq:3}, we then obtain
\begin{align}
\rho ^{\rm nc}(t)=\frac{e^{-\epsilon t }}{2}&\left(\frac{2e^{K} \left(e^{\epsilon t }-1\right) }{e^K\left( e^{2h}+e^h+1 \right)+e^h} \left(  e^{2 h}|++\rangle\langle++| + |--\rangle\langle--| \right)\right.\nonumber \\ 
&+\left( \frac{\left(e^{\epsilon t }-1\right) \cosh \left(\frac{K}{2}\right)}{\cosh \left(\frac{K}{2}\right)+\cosh (h) e^{K/2}}+\cos \omega t +1 \right)|+-\rangle\langle+-|\nonumber \\ 
&+\left( \frac{\left(e^{\epsilon t }-1\right) \cosh \left(\frac{K}{2}\right)}{\cosh \left(\frac{K}{2}\right)+\cosh (h) e^{K/2}}-\cos \omega t +1 \right)|-+\rangle\langle-+|\nonumber \\ &+\left( \frac{e^h \left(e^{\epsilon t }-1\right) \left(e^{K}-1\right)}{e^K\left( e^{2h}+e^h+1 \right)+e^h}-i \sin \omega t  \right)|+-\rangle\langle-+|\nonumber \\ &\left.+\left( \frac{e^h \left(e^{\epsilon t }-1\right) \left(e^{K}-1\right)}{e^K\left( e^{2h}+e^h+1 \right)+e^h}+i \sin \omega t  \right)|-+\rangle\langle+-| \right) ,\label{eq:08nc}
\end{align}
for the non-competing case or
\begin{align}
\rho ^{\rm c}(t)=&\frac{e^{-\epsilon t }}{2}\Biggl(\frac{e^{K/2} \left(e^{\epsilon t }-1\right)}{e^{K/2}+\cosh (L)}\left( |++\rangle\langle++|+|--\rangle\langle--| \right)  \nonumber \\ 
&+\left( \frac{\frac{16 L e^{L+\epsilon t /2} \sinh \left(\frac{\epsilon t }{2}\right) \left(h \sinh (L)+ L \cosh (L)\right)}{2 e^{K/2+L}+e^{2 L}+1}+K^2 \cos \left(\frac{2 L t}{K}\right)-K^2+8 L^2}{4 L^2} \right)\nonumber \\ 
&\times |+-\rangle\langle+-|\nonumber \\ 
&+\left( \frac{\frac{8 L e^{L+\epsilon t /2} \sinh \left(\frac{\epsilon t }{2}\right) \left( L \cosh (L)-h \sinh (L)\right)}{2 e^{K/2+L}+e^{2 L}+1}+K^2 \sin ^2\left(\frac{L t}{K}\right)}{2 L^2} \right)|-+\rangle\langle-+|\nonumber \\ 
&+\frac{K}{4 L} \left(\frac{4h \sin ^2\left(\frac{L t}{K}\right)}{L}+\frac{2 \left(e^{2 L}-1\right) \left(e^{\epsilon t }-1\right)}{2 e^{K/2+L}+e^{2 L}+1}-2 i \sin \left(\frac{2 L t}{K}\right)\right) \nonumber \\ 
&\times |+-\rangle\langle-+|\nonumber \\ 
&+ \frac{K}{4 L} \left(\frac{4h \sin ^2\left(\frac{L t}{K}\right)}{L}+\frac{2 \left(e^{2 L}-1\right) \left(e^{\epsilon t }-1\right)}{2 e^{K/2+L}+e^{2 L}+1}+2 i \sin \left(\frac{2 L t}{K}\right)\right) \nonumber \\ 
&\times |-+\rangle\langle+-| \Biggr), \label{eq:08}
\end{align}
for the competing case, where the scaled time, $\omega t$, and dissipation rate, $\epsilon /\omega $, have been used. In addition we have introduced the definitions $K:=\beta J=\beta \omega \hslash,~h:=\beta g\mu _BH$, and $L:=\sqrt{h^2+K^2/4}$. 

The extended density matrix, $\hat{\rho} ^{\alpha}$, in the TFD doubled Hilbert space was defined in Ref.~\cite{Hashizume} as
\begin{align}
\hat{\rho } ^{\alpha}:=|\Psi \rangle\langle\Psi |,\ |\Psi \rangle:=\left( \rho ^{\alpha}(t) \right) ^{1/2}\sum _{s}|s,\tilde{s}\rangle=\left( \rho ^{\alpha}(t) \right) ^{1/2}\sum _{s}|s\rangle|\tilde{s}\rangle, 
\end{align}
using the ordinary density matrix, $\rho ^{\alpha}(t) $ in Eq.~\eqref{eq:3}. Here, $\left\{ |s\rangle \right\}$ is an orthogonal complete set in the original Hilbert space and $\left\{ |\tilde{s}\rangle \right\}$ the same set in the ancillary Hilbert space of TFD \cite{Suzuki3,Suzuki4}. If the entangled subsystems A and B are being examined, each of the $|s\rangle$ and $|\tilde{s}\rangle$ states is represented as a direct product: $|s_{\text{A}}, s_{\text{B}}\rangle=|s_{\text{A}}\rangle | s_{\text{B}}\rangle$ and $|\tilde{s}_{\text{A}}, \tilde{s}_{\text{B}}\rangle=|\tilde{s}_{\text{A}}\rangle |\tilde{s}_{\text{B}}\rangle$, respectively. 

\begin{figure}[hbpt]
\begin{center}
\unitlength 1mm
\begin{picture}(150,80)
\put(0,45){
\scalebox{0.65}{
\put(0,0){\includegraphics{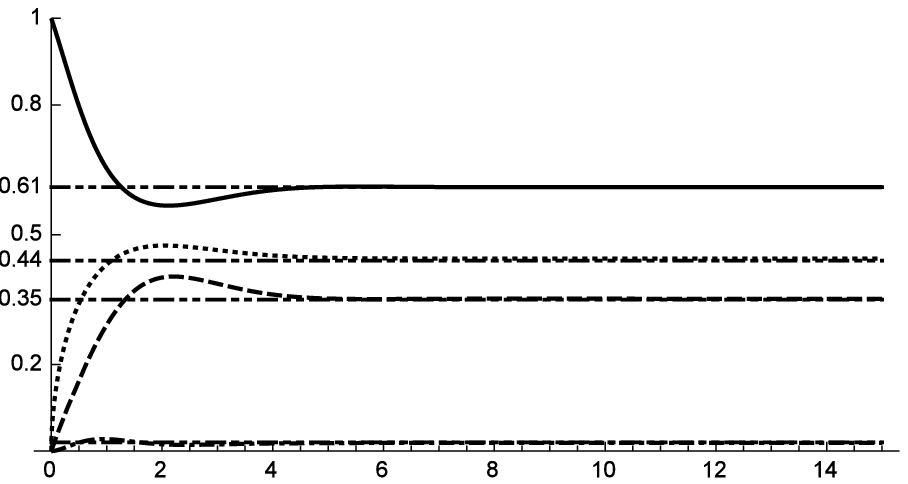}}
\put(94,.5){$\omega t$}
\put(7,44){$b_{\text{d1}}^{\rm nc}$}
\put(7,25){$b_{\text{cf}}^{\rm nc}$}
\put(10,10){$b_{\rm d2}^{\rm nc}$}
\put(18,6){$b_{\rm qe}^{\rm nc}$}
\put(47,-7){\scalebox{1.5}{(a)}}
}
}
\put(71,45){
\scalebox{0.65}{
\put(0,0){\includegraphics{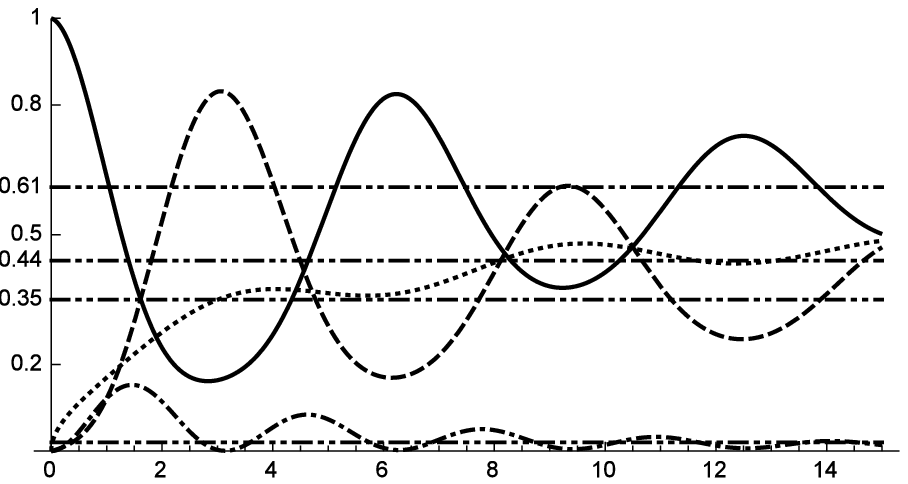}}
\put(94,.5){$\omega t$}
\put(8,44){$b_{\text{d1}}^{\rm nc}$}
\put(18,21){$b_{\text{cf}}^{\rm nc}$}
\put(25,38){$b_{\rm d2}^{\rm nc}$}
\put(28,9){$b_{\rm qe}^{\rm nc}$}
\put(47,-7){\scalebox{1.5}{(b)}}
}
}
\put(0,0){
\scalebox{0.65}{
\put(0,0){\includegraphics{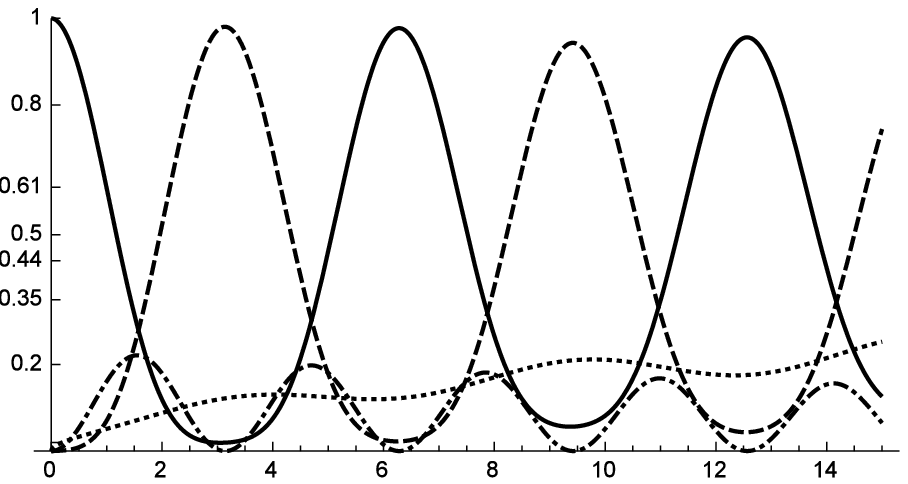}}
\put(94,.5){$\omega t$}
\put(8,44){$b_{\text{d1}}^{\rm nc}$}
\put(20,10.5){$b_{\text{cf}}^{\rm nc}$}
\put(25,44){$b_{\rm d2}^{\rm nc}$}
\put(7,13){$b_{\rm qe}^{\rm nc}$}
\put(47,-7){\scalebox{1.5}{(c)}}
}
}
\put(71,0){
\scalebox{0.65}{
\put(0,0){\includegraphics{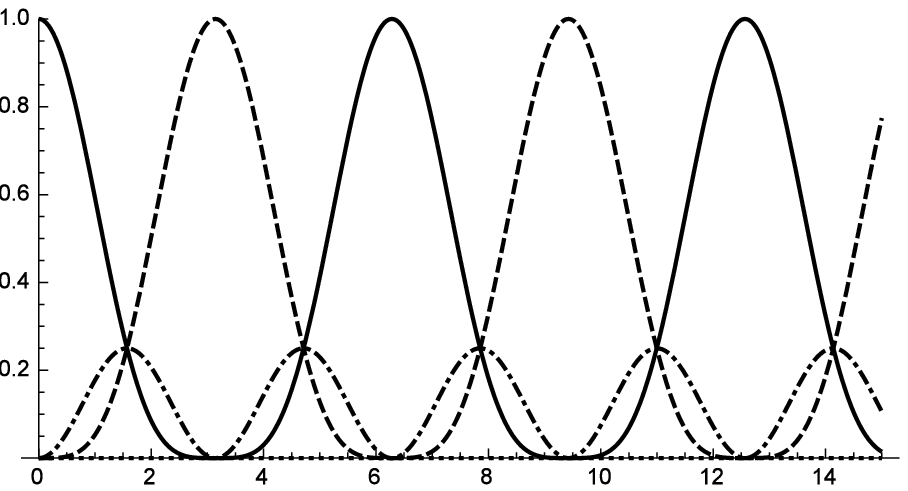}}
\put(94,.5){$\omega t$}
\put(7,44){$b_{\text{d1}}^{\rm nc}$}
\put(25,44){$b_{\rm d2}^{\rm nc}$}
\put(17,11){$b_{\rm qe}^{\rm nc}$}
\put(47,-7){\scalebox{1.5}{(d)}}
}
}
\end{picture}
\end{center}
\caption{Time dependence of the matrix elements, $b_{\text{d1}}^{\rm nc}(\mbox{\bf -\hspace{-.4mm}\bf -\hspace{-.5mm}\bf -\hspace{-.4mm}\bf -\hspace{-.4mm}\bf -}), b_{\text{d2}}^{\rm nc}(\mbox{\bf -\hspace{0mm}\bf -\hspace{0mm}\bf -}), b_{\text{cf}}^{\rm nc}(\mbox{\boldmath$\cdot \hspace{-1mm}\cdot \hspace{-1mm}\cdot $})$, and $b_{\text{qe}}^{\rm nc}(\mbox{\bf -\boldmath $\cdot $\hspace{.5mm}\bf -})$, in dissipative and non-dissipative systems with $K^{-1}=(\beta J)^{-1}=0.7$ and $h=\beta g\mu_{B}H=3/7$, for the non-competing cases. Parts (a), (b), and (c) show cases with a scaled dissipation rate $\epsilon /\omega =1,~0.1$, and $0.01$, respectively. The dot-dot-dashed lines (\mbox{\bf -\boldmath $\cdot \cdot $\hspace{.5mm}\bf -}) in parts (a) and (b) represent the asymptotes of the $b_{\text{d1}}^{\rm nc}, b_{\text{cf}}^{\rm nc}, b_{\text{d2}}^{\rm nc}$, and $b_{\text{qe}}^{\rm nc}$ curves, respectively. In part (d), which is the non-dissipation case ($\epsilon =0$), $b_{\text{cf}}^{\rm nc}$ vanishes identically. On the other hand, $b_{\text{d1}}^{\rm nc}$, $b_{\text{d2}}^{\rm nc}$, and $b_{\text{qe}}^{\rm nc}$ show a kind of ``classical-quantum crossover'' oscillation.}\label{fig:01nc}
\end{figure}

\begin{figure}[hbpt]
\begin{center}
\unitlength 1mm
\begin{picture}(150,90)
\put(0,53){
\scalebox{0.65}{
\put(0,0){\includegraphics{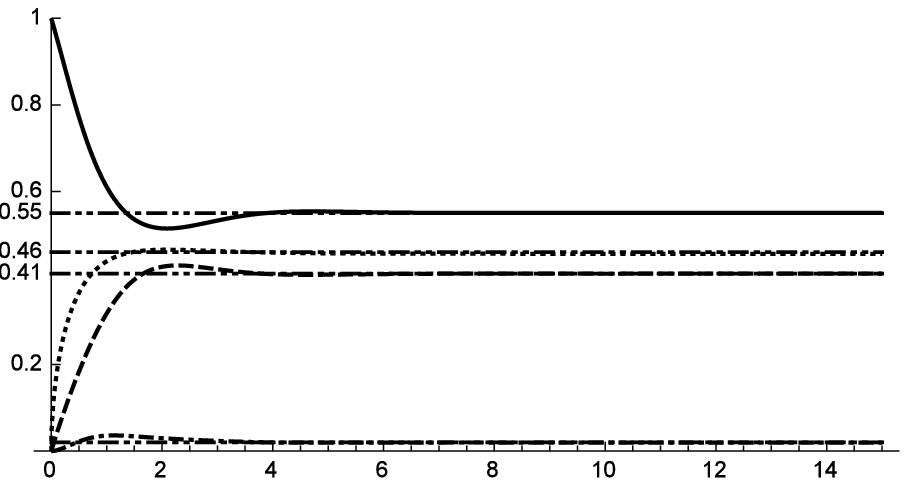}}
\put(94,.5){$\omega t$}
\put(7,44){$b_{\text{d1}}^{\rm c}$}
\put(7,25){$b_{\text{cf}}^{\rm c}$}
\put(9,10){$b_{\rm d2}^{\rm c}$}
\put(18,7){$b_{\rm qe}^{\rm c}$}
\put(47,-7){\scalebox{1.5}{(a)}}
}
}
\put(72,53){
\scalebox{0.65}{
\put(0,0){\includegraphics{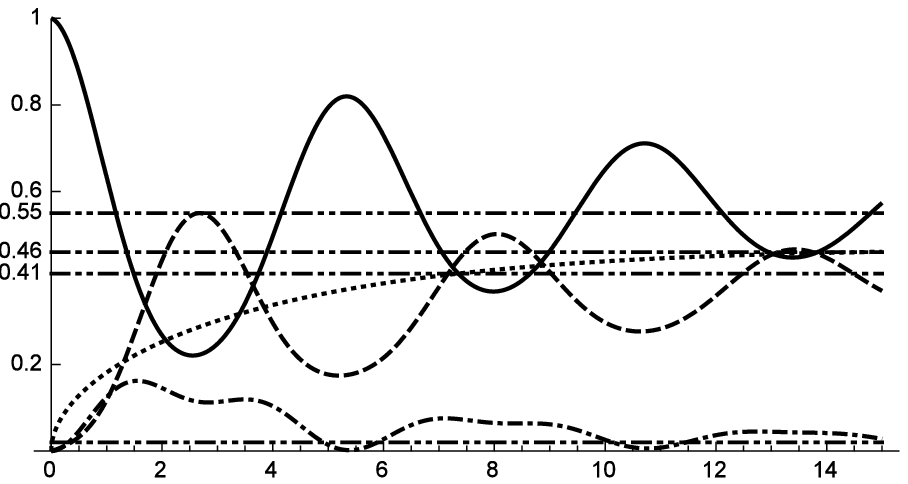}}
\put(94,.5){$\omega t$}
\put(7.5,44){$b_{\text{d1}}^{\rm c}$}
\put(18,18){$b_{\text{cf}}^{\rm c}$}
\put(15,29){$b_{\rm d2}^{\rm c}$}
\put(14,6){$b_{\rm qe}^{\rm c}$}
\put(47,-7){\scalebox{1.5}{(b)}}
}
}
\put(0,0){
\scalebox{0.65}{
\put(0,0){\includegraphics{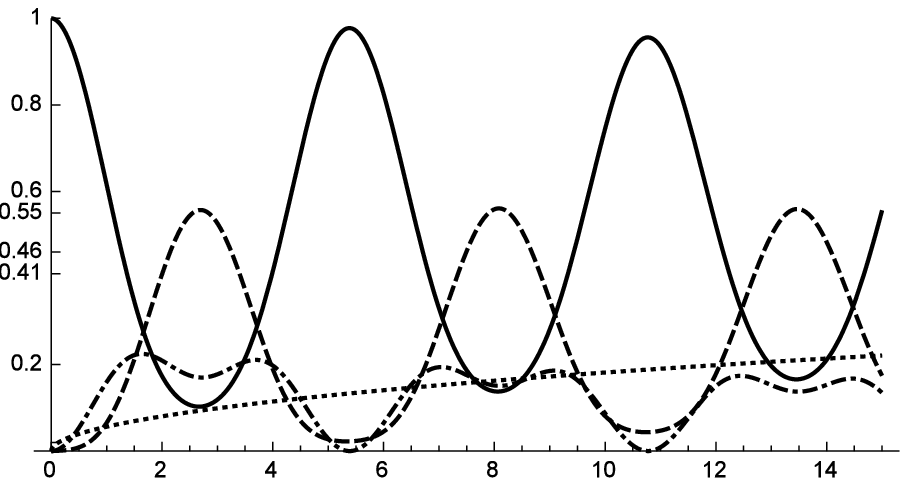}}
\put(94,.5){$\omega t$}
\put(8,44){$b_{\text{d1}}^{\rm c}$}
\put(34,11){$b_{\text{cf}}^{\rm c}$}
\put(15,29){$b_{\rm d2}^{\rm c}$}
\put(7,13){$b_{\rm qe}^{\rm c}$}
\put(47,-7){\scalebox{1.5}{(c)}}
}
}
\put(72,0){
\scalebox{0.65}{
\put(0,0){\includegraphics{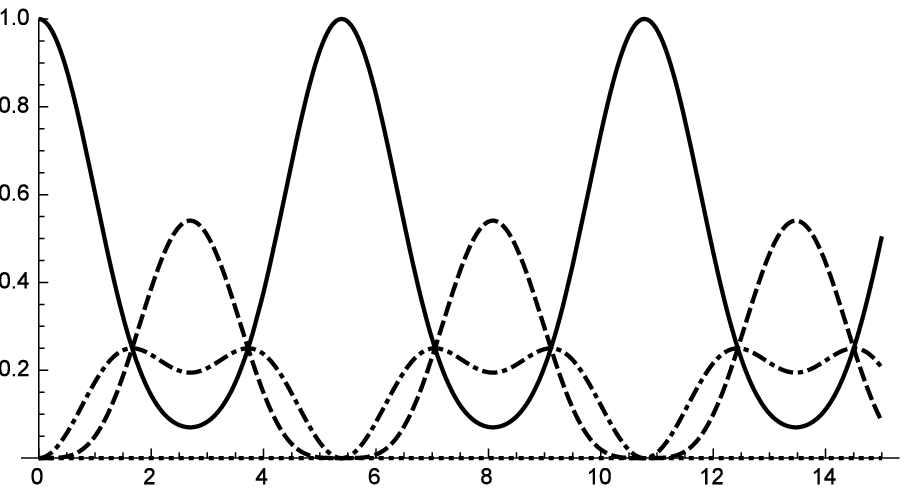}}
\put(94,.5){$\omega t$}
\put(8,44){$b_{\text{d1}}^{\rm c}$}
\put(15,29){$b_{\rm d2}^{\rm c}$}
\put(6,14){$b_{\rm qe}^{\rm c}$}
\put(47,-7){\scalebox{1.5}{(d)}}
}
}
\end{picture}
\end{center}
\caption{Time dependence of the matrix elements, $b_{\text{d1}}^{\rm c}(\mbox{\bf -\hspace{-.4mm}\bf -\hspace{-.5mm}\bf -\hspace{-.4mm}\bf -\hspace{-.4mm}\bf -}), b_{\text{d2}}^{\rm c}(\mbox{\bf -\hspace{0mm}\bf -\hspace{0mm}\bf -}), b_{\text{cf}}^{\rm c}(\mbox{\boldmath$\cdot \hspace{-1mm}\cdot \hspace{-1mm}\cdot $})$, and $b_{\text{qe}}^{\rm c}(\mbox{\bf -\boldmath $\cdot $\hspace{.5mm}\bf -})$, in dissipative and non-dissipative systems with $K^{-1}=(\beta J)^{-1}=0.7$ and $h=\beta g\mu_{B}H=3/7$ for the competing cases. Parts (a), (b), and (c) show cases with a scaled dissipation rate of $\epsilon /\omega =1,~0.1$, and $0.01$, respectively. The dot-dot-dashed lines (\mbox{\bf -\boldmath $\cdot \cdot $\hspace{.5mm}\bf -}) in parts (a) and (b) represent the asymptotes of the $b_{\text{d1}}^{\rm c}, b_{\text{cf}}^{\rm c}, b_{\text{d2}}^{\rm c}$, and $b_{\text{qe}}^{\rm c}$ curves, respectively. In part (d), the non-dissipation case ($\epsilon =0$), $b_{\text{cf}}^{\rm c}$ vanishes identically, while the $b_{\text{qe}}^{\rm c}$ curve displays a ``twin-peaks'' oscillation.}\label{fig:01}
\end{figure}

According to the matrix algebra of $\rho ^{\alpha}(t)$ developed in Ref.~\cite{Hashizume}, we then obtain the extended density matrix,
\begin{align}
\hat{\rho}_{\text{A}}^{\alpha} &=b_{\text{d1}}^{\alpha}|+\rangle\langle+||\tilde{+}\rangle\langle\tilde{+}| +b_{\text{d2}}^{\alpha}|-\rangle\langle-||\tilde{-}\rangle\langle\tilde{-}|
\nonumber \\ &+b_{\text{cf}}^{\alpha}\left(
|+\rangle\langle-||\tilde{+}\rangle\langle\tilde{-}|+|-\rangle\langle+||\tilde{-}\rangle\langle\tilde{+}| \right)\nonumber \\
&+b_{\text{qe}}^{\alpha}\left( |+\rangle\langle+||\tilde{-}\rangle\langle\tilde{-}|+|-\rangle\langle-||\tilde{+}\rangle\langle\tilde{+}|
\right), \label{eq:17}
\end{align}
where the matrix elements, $b_{\text{d1}}^{\alpha}, b_{\text{d2}}^{\alpha}, b_{\text{cf}}^{\alpha}$, and $b_{\text{qe}}^{\alpha}$, are obtained as analytic functions of $H,~\epsilon,~t$, and $\beta$, respectively. They correspond to the two diagonal elements (d1 and d2), classical and thermal fluctuations (cf), and quantum entanglement (qe) of $\hat{\rho}_{\text{A}}^{\alpha}$, respectively.  Their expressions are so complicated that we show only the numerical results in Fig.~\ref{fig:01nc} for the non-competing cases and in Fig.~\ref{fig:01} for the competing cases. The asymptotic behaviors of  $b_{\text{d1}}^{\alpha}, b_{\text{d2}}^{\alpha}, b_{\text{cf}}^{\alpha}$, and $b_{\text{qe}}^{\alpha}$ at $t\to\infty$ correspond to those of the equilibrium systems as follows:
\begin{align}
\lim_{t\to\infty}b_{\text{d1}}^{\rm nc}&=\frac{4 e^{h+K}+\left(e^{K/2}+1\right)^2}{4 e^K (2 \cosh (h)+1)+4},\nonumber \\
\lim_{t\to\infty}b_{\text{d2}}^{\rm nc}&=\frac{\left(e^h+4\right) e^K+2 e^{h+\frac{K}{2}}+e^h}{4 \left(\left(e^h+e^{2 h}+1\right) e^K+e^h\right)},\nonumber \\
\lim_{t\to\infty}b_{\text{cf}}^{\rm nc}&=\frac{\left(e^h+1\right) \left(e^{K/2}+1\right) e^{\frac{h+K}{2}}}{2 \left(\left(e^h+e^{2 h}+1\right) e^K+e^h\right)},\nonumber \\
\mbox{and }\lim_{t\to\infty}b_{\text{qe}}^{\rm nc}&=\frac{\left(e^{K/2}-1\right)^2}{4 e^K (2 \cosh (h)+1)+4}
\label{eq:17a}
\end{align}
and
\begin{align}
\lim_{t\to\infty}b_{\text{d1}}^{\rm c}&=\frac{e^{-L}}{32 L^2 \left(e^{K/2}+\cosh (L)\right)} \Bigl(4 L \Bigl(\left(e^{2 L}-1\right) \sqrt{4 L^2-K^2}\nonumber \\
&\hspace{5mm}+2 L \left(2 e^{\frac{K}{2}+L}+e^{2 L}+1\right)\Bigr)-K^2 \left(e^L-1\right)^2\Bigr),\nonumber \\
\lim_{t\to\infty}b_{\text{d2}}^{\rm c}&=\frac{e^{-L}}{32 L^2 \left(e^{K/2}+\cosh (L)\right)} \Bigl(4 L \Bigl(2 L \left(2 e^{\frac{K}{2}+L}+e^{2 L}+1\right)\nonumber \\
&\hspace{5mm}-\left(e^{2 L}-1\right) \sqrt{4 L^2-K^2}\Bigr)-K^2 \left(e^L-1\right)^2\Bigr),\nonumber \\
\lim_{t\to\infty}b_{\text{cf}}^{\rm c}&=\frac{\left(e^L+1\right) e^{\frac{1}{4} (K+2 L)}}{2 e^{\frac{K}{2}+L}+e^{2 L}+1},\nonumber \\
\mbox{and }\lim_{t\to\infty}b_{\text{qe}}^{\rm c}&=\frac{K^2 \left(e^L-1\right)^2}{16 L^2 \left(2 e^{\frac{K}{2}+L}+e^{2 L}+1\right)},
\label{eq:17b}
\end{align}
respectively. The asymptotic values with $K^{-1}=(\beta J)^{-1}=0.7$ and $h=\beta g\mu_{B}H=3/7$, which are obtained from Eqs.~\eqref{eq:17a} and \eqref{eq:17b}, are shown in Figs.~1 and 2. It is worth mentioning that the findings of Figs.~1(a), (b), and (c), and Figs.~2(a), (b), and (c) may be in harmony with the relaxation processes of entanglement by dissipation.

As can be seen from Fig.~1(d), which shows the non-dissipation case $(\epsilon=0)$, $b_{\text{cf}}^{\alpha}$ vanishes identically. On the other hand, $b_{\text{d1}}^{\rm nc}$, $b_{\text{d2}}^{\rm nc}$, and $b_{\text{qe}}^{\rm nc}$ show a kind of ``classical-quantum crossover'' oscillation in Ref.~\cite{Hashizume}. However, as can be seen from Fig.~2(d), which shows the non-dissipation case $(\epsilon=0)$, $b_{\text{cf}}^{\alpha}$ vanishes identically, while the $b_{\text{qe}}^{\rm c}$ curve displays a ``twin-peaks'' oscillation, which is a new type.

In Ref.~\cite{Hashizume}, the following results were obtained, illustrating the equilibrium case: i) In the non-competing systems (for zero temperature), the external field, $H$, breaks the spin inversion symmetry and the parameter $b_{\text{d1}}^{\rm nc}=1$, because the quantum entanglement $b_{\text{qe}}^{\rm nc}=0$. ii) In the competing system, the level splitting creates a non-zero finite entanglement, $b_{\text{qe}}^{\rm c}\ne 0$, for zero temperature, even in a finite external field, $H$. iii) The parameter, $b_{\text{d1}}^{\rm c}$ , which expresses the probability weight of the up-state, is smaller than the maximum value, $1.0$. According to these observations, Ref.~\cite{Hashizume} demonstrated a typical example of entanglement caused by the external field.

Therefore, also in the present analysis it seems reasonable to conclude that the competing ``twin-peaks'' oscillatory behavior is a consequence of the partial recovery of the spin inversion symmetry even in the dissipative dynamics.

\newpage
	
\section{Extended entanglement entropies of non-equilibrium finite-spin systems with external fields \label{sec:3}}
\begin{figure}[hbpt]
\begin{center}
\unitlength 1mm
\begin{picture}(150,90)
\put(0,53){
\scalebox{0.65}{
\put(0,0){\includegraphics{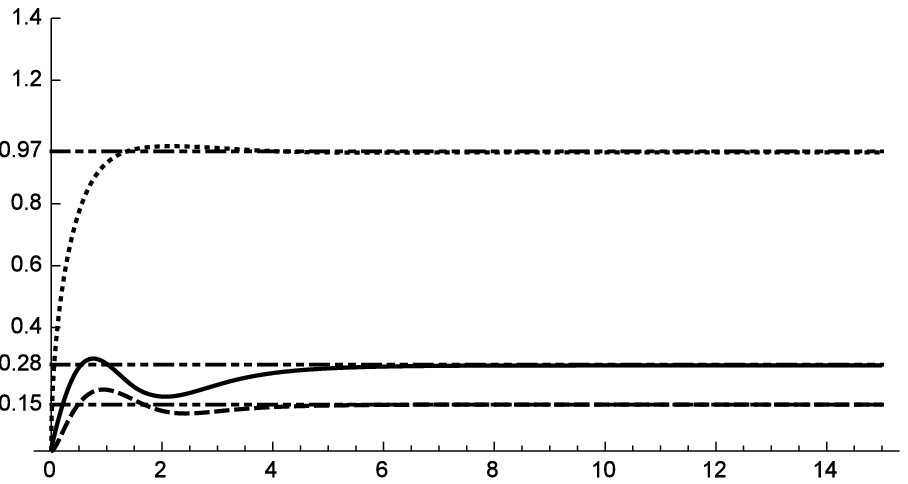}}
\put(92,.5){$\omega t$}
\put(91,11){{$\hat{S}^{\rm nc}$}}
\put(91,6){{$\hat{S}_{\text{qe}}^{\rm nc}$}}
\put(91,33){{$C^{\rm nc}$}}
\put(47,-7){\scalebox{1.5}{(a)}}
}
}
\put(71,53){
\scalebox{0.65}{
\put(0,0){\includegraphics{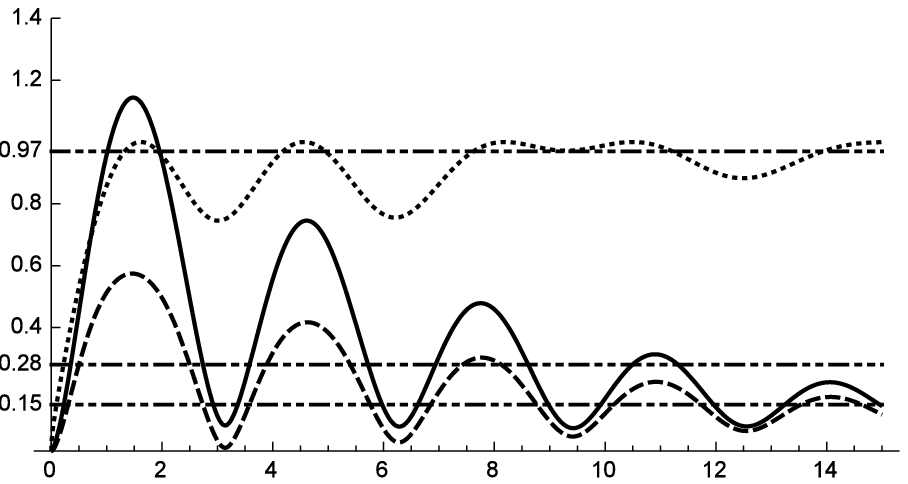}}
\put(92,.5){$\omega t$}
\put(12,40){$\hat{S}^{\rm nc}$}
\put(12,23){$\hat{S}_{\text{qe}}^{\rm nc}$}
\put(21,23){$C^{\rm nc}$}
\put(47,-7){\scalebox{1.5}{(b)}}
}
}
\put(0,0){
\scalebox{0.65}{
\put(0,0){\includegraphics{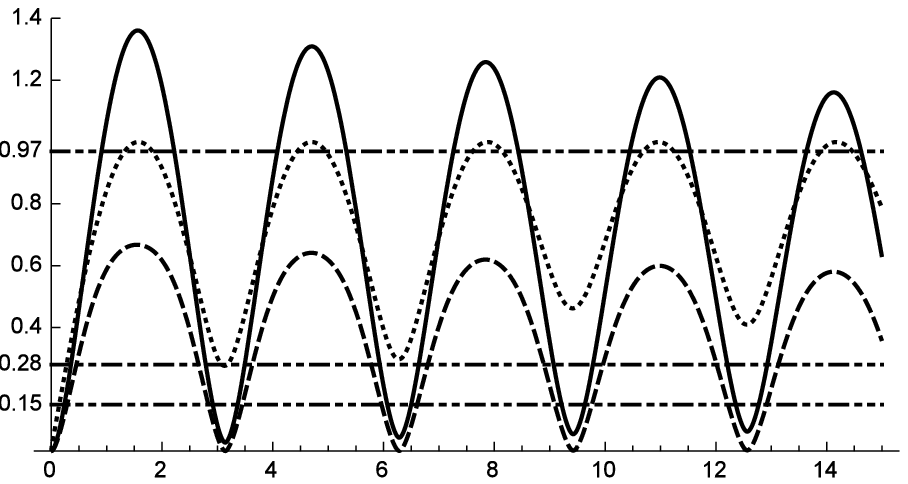}}
\put(92,.5){$\omega t$}
\put(16,45){$\hat{S}^{\rm nc}$}
\put(12,19){$\hat{S}_{\text{qe}}^{\rm nc}$}
\put(11,27){$C^{\rm nc}$}
\put(47,-7){\scalebox{1.5}{(c)}}
}
}
\put(71,0){
\scalebox{0.65}{
\put(1.5,0){\includegraphics{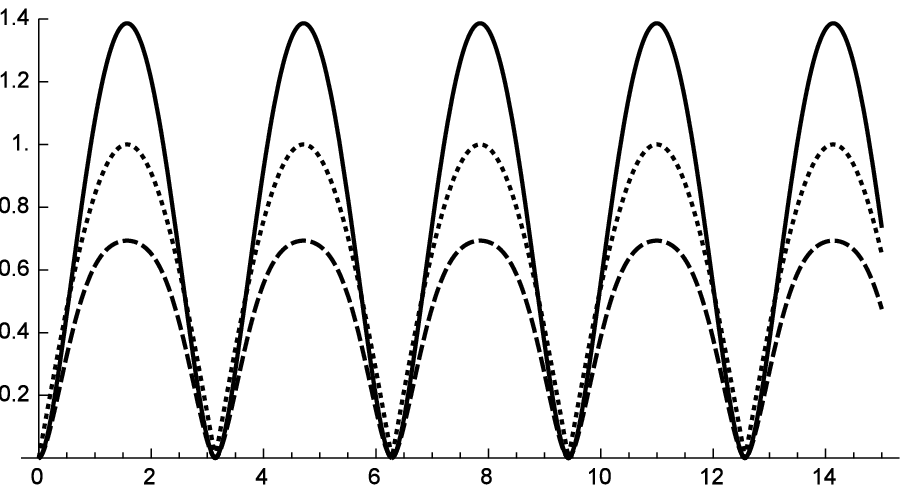}}
\put(92,.5){$\omega t$}
\put(17,45){$\hat{S}^{\rm nc}$}
\put(12,20){$\hat{S}_{\text{qe}}^{\rm nc}$}
\put(11.5,36){$C^{\rm nc}$}
\put(47,-7){\scalebox{1.5}{(d)}}
}
}
\end{picture}
\end{center}
\caption{Time dependence of $\hat{S}^{\rm nc}(\mbox{\bf -\hspace{-.4mm}\bf -\hspace{-.5mm}\bf -\hspace{-.4mm}\bf -\hspace{-.4mm}\bf -})$, $\hat{S}_{\text{qe}}^{\rm nc}(\mbox{\bf -\hspace{0mm}\bf -\hspace{0mm}\bf -})$, and  $C^{\rm nc}(\mbox{\boldmath$\cdot \hspace{-1mm}\cdot \hspace{-1mm}\cdot $})$ 
in dissipative and non-dissipative systems with $K^{-1}=(\beta J)^{-1}=0.7$ and $h=\beta g\mu_{B}H=3/7$ for the non-competing cases. Parts (a), (b), and (c) show cases with scaled dissipation rates of $\epsilon /\omega =1,~0.1$, and $0.01$, respectively. The dot-dot-dashed lines (\mbox{\bf -\boldmath $\cdot \cdot $\hspace{.5mm}\bf -}) in parts (a), (b), and (c) represent the asymptotes of the $\hat{S}^{\rm nc}$ and $\hat{S}_{\rm qe}^{\rm nc}$ curves. In part (d), which is the non-dissipation case ($\epsilon =0$), all the curves have the same phase. In these and all subsequent plots, $k_{\text{B}} = 1$.}\label{fig:02nc}
\end{figure}

\begin{figure}[hbpt]
\begin{center}
\unitlength 1mm
\begin{picture}(150,90)
\put(0,53){
\scalebox{0.65}{
\put(0,0){\includegraphics{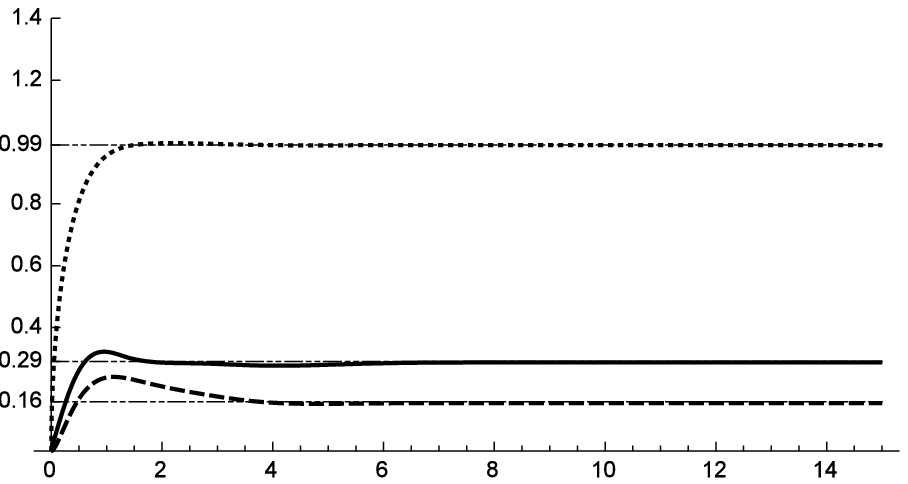}}
\put(91,11){{$\hat{S}^{\rm c}$}}
\put(91,7){{$\hat{S}_{\text{qe}}^{\rm c}$}}
\put(91,34){{$C^{\rm c}$}}
\put(92,.5){$\omega t$}
\put(47,-7){\scalebox{1.5}{(a)}}
}
}
\put(71,53){
\scalebox{0.65}{
\put(0,0){\includegraphics{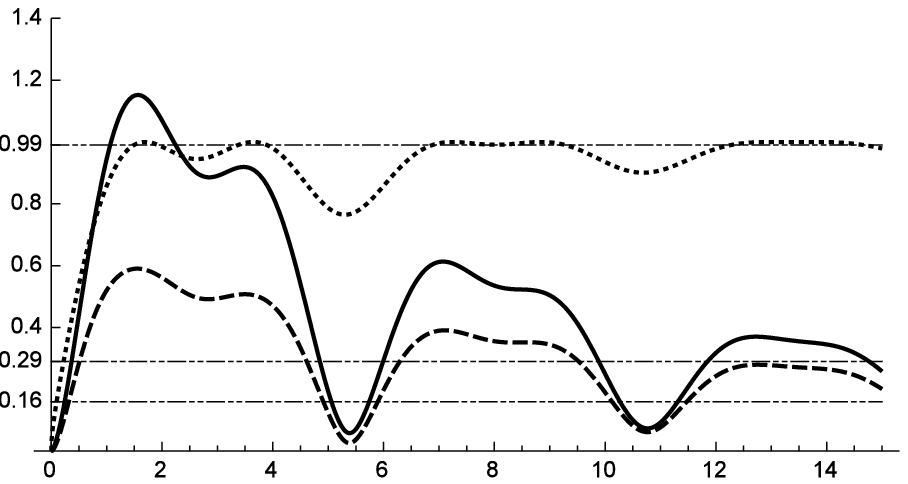}}
\put(92,.5){$\omega t$}
\put(12,41){$\hat{S}^{\rm c}$}
\put(12,24){$\hat{S}_{\text{qe}}^{\rm c}$}
\put(36,24){$C^{\rm c}$}
\put(47,-7){\scalebox{1.5}{(b)}}
}
}
\put(0,0){
\scalebox{0.65}{
\put(0,0){\includegraphics{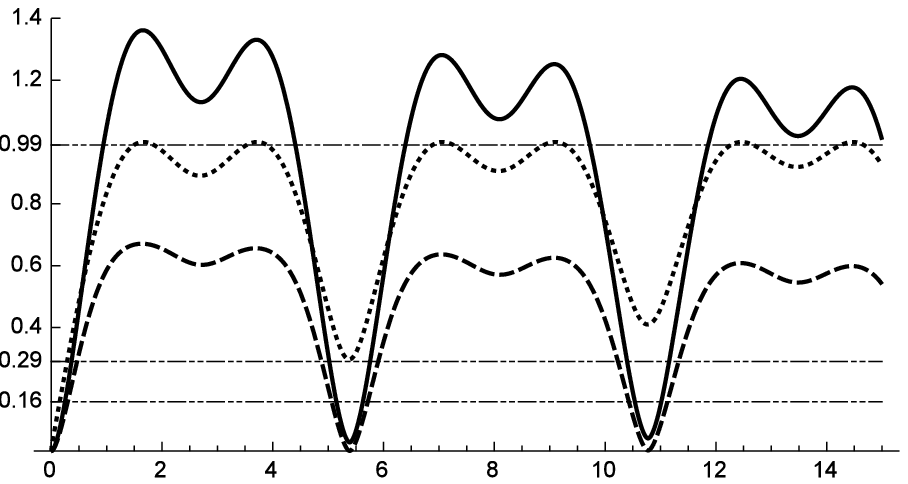}}
\put(92,.5){$\omega t$}
\put(17,45){$\hat{S}^{\rm c}$}
\put(12,26){$\hat{S}_{\text{qe}}^{\rm c}$}
\put(13,35){$C^{\rm c}$}
\put(47,-7){\scalebox{1.5}{(c)}}
}
}
\put(71,0){
\scalebox{0.65}{
\put(1.5,0){\includegraphics{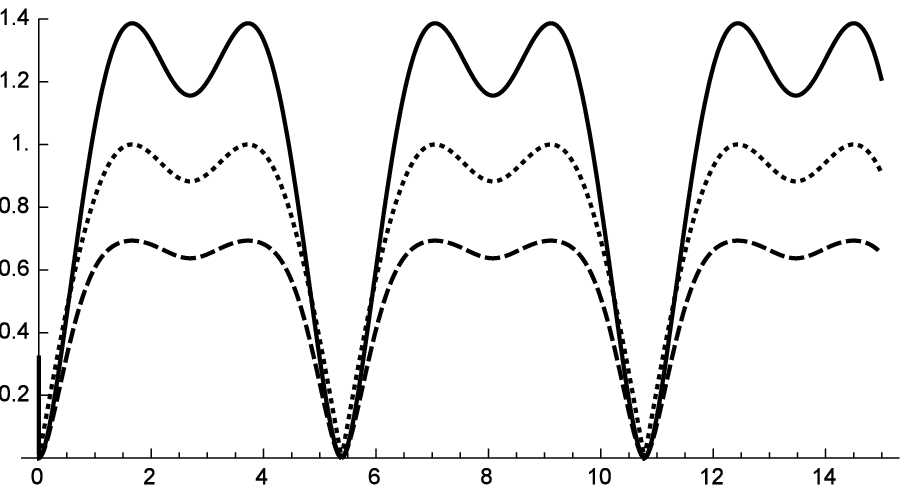}}
\put(93,.5){$\omega t$}
\put(17,46){$\hat{S}^{\rm c}$}
\put(13,27){$\hat{S}_{\text{qe}}^{\rm c}$}
\put(16,35){$C^{\rm c}$}
\put(47,-7){\scalebox{1.5}{(d)}}
}
}
\end{picture}
\end{center}
\caption{Time dependence of $\hat{S}^{\rm c}(\mbox{\bf -\hspace{-.4mm}\bf -\hspace{-.5mm}\bf -\hspace{-.4mm}\bf -\hspace{-.4mm}\bf -})$, $\hat{S}_{\text{qe}}^{\rm c}(\mbox{\bf -\hspace{0mm}\bf -\hspace{0mm}\bf -})$, and  $C^{\rm c}(\mbox{\boldmath$\cdot \hspace{-1mm}\cdot \hspace{-1mm}\cdot $})$ in dissipative and non-dissipative systems with $K^{-1}=(\beta J)^{-1}=0.7$ and $h=\beta g\mu_{B}H=3/7$ for the competing cases. Parts (a), (b), and (c) show cases with scaled dissipation rates of $\epsilon /\omega =1,~0.1$, and $0.01$, respectively. The dot-dot-dashed lines (\mbox{\bf -\boldmath $\cdot \cdot $\hspace{.5mm}\bf -}) in parts (a), (b), and (c) represent the asymptotes of the $\hat{S}^{\rm c}$ and $\hat{S}_{\rm qe}^{\rm c}$ curves, respectively.  In part (d), which is the non-dissipation case ($\epsilon =0$), all the curves have the same phase and show a kind of ``twin-peaks'' oscillation.}\label{fig:02}
\end{figure}

The extended entanglement entropy is defined as~\cite{Hashizume}
\begin{align}
\hat{S}^{\alpha }&:=-k_{\text{B}}\mathrm{Tr}_{\text{A}}\left[ \hat{\rho}_{\text{A}}^{\alpha }\log \hat{\rho}_{\text{A}}^{\alpha } \right] ,
\label{eq:22}
\end{align}
using $\hat{\rho }_{\text{A}}^{\alpha}$ in Eq.~\eqref{eq:17}. Inserting Eq.~\eqref{eq:17} into Eq.~\eqref{eq:22} and subsequent simplification eventually yields
\begin{equation}
\hat{S}^{\alpha }=\hat{S}_{\rm cl}^{\alpha }+\hat{S}_{\rm qe}^{\alpha },
\label{eq:23}
\end{equation}
where
\begin{align}
\hat{S}_{\rm cl}^{\alpha }&:=-k_{\text{B}}\left( \sqrt{4\left( b_{\text{cf}}^{\alpha } \right) ^2+\left(b_{\text{d1}}^{\alpha }-b_{\text{d2}}^{\alpha }\right)^2} ~\mathrm{arccoth}\,
\frac{b_{\text{d1}}^{\alpha }+b_{\text{d2}}^{\alpha }}{\sqrt{4\left( b_{\text{cf}}^{\alpha } \right) ^2+\left(b_{\text{d1}}^{\alpha }-b_{\text{d2}}^{\alpha }\right)^2}}\right.\notag \\
&\hspace{3mm}\left. +\frac{b_{\text{d1}}^{\alpha }+b_{\text{d2}}^{\alpha }}{2}\log \left(b_{\text{d1}}^{\alpha }b_{\text{d2}}^{\alpha }-\left( b_{\text{cf}}^{\alpha } \right) ^2\right)\right),
\label{eq:23a}
\end{align}
and
\begin{align}
\hat{S}_{\rm qe}^{\alpha }:=-2k_{\text{B}}b_{\text{qe}}^{\alpha }\log b_{\text{qe}}^{\alpha },
\label{eq:23b}
\end{align}
respectively.
In Eqs.~\eqref{eq:23}, \eqref{eq:23a}, and \eqref{eq:23b}, the expressions for $\hat{S}^{\alpha }$, the classical and thermal fluctuation parts, $\hat{S}_{\rm cl}^{\alpha }$, and the quantum entanglement part, $\hat{S}_{\rm qe}^{\alpha }$, also involve analytic functions of $ t, \beta, \epsilon$, and $\omega$. However, these expressions are quite complicated.
Therefore, we show the numerical behavior of $\hat{S}^{\alpha },~\hat{S}_{\text{qe}}^{\alpha }$, and $b_{\text{qe}}^{\alpha }$ for a few scenarios with $K^{-1}=(\beta J)^{-1}=0.7$ and $h=\beta g\mu_{B}H=3/7$ in Fig.~\ref{fig:02nc} for the non-competing case and in Fig.~\ref{fig:02} for the competing case.
The asymptotic behaviors of $\hat{S}^{\alpha }$ and $\hat{S}_{\rm qe}^{\alpha }$ as $t\to\infty$ correspond to those of equilibrium systems obtained by inserting Eqs.~\eqref{eq:17a} and \eqref{eq:17b} into Eqs.~\eqref{eq:23}, \eqref{eq:23a}, and \eqref{eq:23b}. The asymptotic values of $\hat{S}^{\alpha }$ and $\hat{S}_{\rm qe}^{\alpha }$ with $K^{-1}=(\beta J)^{-1}=0.7$ and $h=\beta g\mu_{B}H=3/7$, obtained from Eqs.~\eqref{eq:23}, \eqref{eq:23a}, and \eqref{eq:23b}, are displayed in Figs.~3 and 4. The data in Figs.~3(a), (b), and (c), and Figs.~4(a), (b), and (c) are in good agreement with the relaxation phenomena of entanglement by dissipation.
 
In the non-dissipative systems, $(\epsilon =0)$, $\hat{S}_{\rm qe}^{\alpha}$ in Eq.~\eqref{eq:23b} reduces to
\begin{align}
\hat{S}_{\rm qe}^{\rm nc}= \frac{k_{\text{B}}}{2} \sin ^2 t\cdot \log \left( 4\csc ^2 t \right),
\label{eq:25nc}
\end{align}
for the non-competing case and
\begin{align}
\hat{S}_{\rm qe}^{\rm c}=&\frac{k_{\text{B}}\left(\cos \left(\sqrt{4 h^2+1} t\right)+8 h^2+1\right) }{\left(4 h^2+1\right)^2}\sin ^2\left(\frac{1}{2} \sqrt{4 h^2+1} t\right) \nonumber \\
&\times \left(\log \left(\frac{2 \csc ^2\left(\frac{1}{2} \sqrt{4 h^2+1} t\right)}{\cos \left(\sqrt{4 h^2+1} t\right)+8 h^2+1}\right)+2 \log (4 h^2+1) \right)
\label{eq:25}
\end{align}
for the competing case. As can be seen from Eqs.~\eqref{eq:25nc} and \eqref{eq:25}, both the $K$ and $H$ dependences of $\hat{S}_{\rm qe}^{\rm nc}$ at $\epsilon =0$ disappear and the $K$ dependence of $\hat{S}_{\rm qe}^{\rm c}$ at $\epsilon =0$ disappears. The time dependence of $\hat{S}^{\alpha }$ and $\hat{S}_{\rm qe}^{\alpha }$ at $\epsilon =0$ is shown in Fig.~\ref{fig:02nc}(d) for the non-competing case and in Fig.~\ref{fig:02}(d) for the competing case. It is apparent from these figures that the curves ($\hat{S}^{\alpha }$ and $\hat{S}_{\text{qe}}^{\alpha }$), showing entanglement, have the same phase. However, their amplitudes differ. 
Especially, the $H$ dependences of $\hat{S}_{\rm qe}^{\rm c}$ at $\epsilon =0$ are displayed in Fig.~5. 
From this observation we infer that the ``twin-peaks'' oscillation is an example of quantum entanglement caused by level splitting.

Next, we consider the concurrence of non-equilibrium finite-spin systems with external fields. The ordinary entanglement entropy, $S^{\alpha }$, is defined by
\begin{align}
S^{\alpha }&:=-k_{\text{B}}{\text{Tr}}_{\text{A}}\left[ \rho ^{\alpha }_{\text{A}}\log \rho ^{\alpha }_{\text{A}} \right], ~\mbox{and }~\rho ^{\alpha }_{\text{A}}:={\text{Tr}}_{\text{B}}\rho ^{\alpha }(t). \label{eq:09}
\end{align}
The insertion of Eq.~\eqref{eq:08} into Eq.~\eqref{eq:09} yields
\begin{align}
S^{\alpha}=&-k_{\text{B}}\left( \frac{1+\sqrt{1-\left( C^{\alpha} \right) ^2}}{2}\log \left( \frac{1+\sqrt{1-\left( C^{\alpha} \right) ^2}}{2} \right) \right.\nonumber \\
&\left.+ \frac{1-\sqrt{1-\left( C^{\alpha} \right) ^2}}{2}\log \left( \frac{1-\sqrt{1-\left( C^{\alpha} \right) ^2}}{2} \right) \right) ,\label{eq:11}
\end{align}
where ${C^{\alpha}}$ is the concurrence \cite{Wootters} for the competing or non-competing cases.
The time dependence of $C^{\alpha }$ is displayed in Fig.~\ref{fig:02nc} for the non-competitive case and in Fig.~\ref{fig:02} for the competitive case (in units of $k_{\text{B}}=1$). As can clearly be seen from Figs.~\ref{fig:02nc} and \ref{fig:02}, $C^{\alpha}$ includes not only the contribution from the quantum entanglement, but also the contributions from the classical and thermal fluctuations in the non-equilibrium case. However, this fact is not manifest in the above expressions for $S^{\alpha}$ and $C^{\alpha}$.

%

\begin{figure}[hbpt]
\begin{center}
\scalebox{.8}{\includegraphics{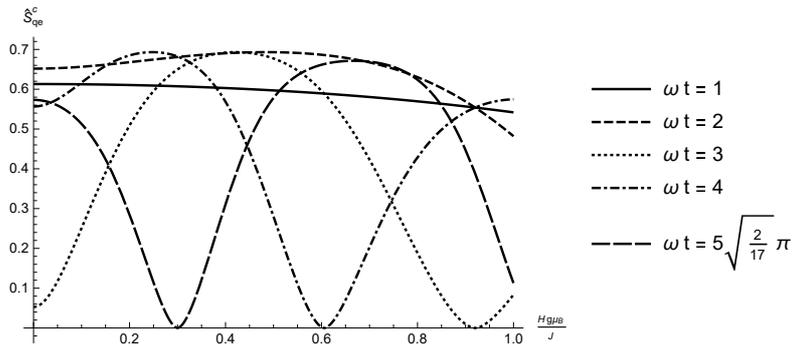}}
\end{center}
\caption{$H$ dependences of $\hat{S}_{\rm qe}^{\rm c}$ in non-dissipative system $(\epsilon =0)$ at $\omega t=1,~\cdots ,~5\sqrt{\frac{2}{17}}\,\pi$. Here, $\omega t=5\sqrt{\frac{2}{17}}\,\pi=5.38779\cdots$ is the cycle time of the oscillation in Fig.\,2(d)}\label{fig:05}
\end{figure}

\newpage

\section{Discussion and Conclusions \label{sec:4}}
In this communication, we have examined the extended density matrix and the extended entanglement entropies of non-equilibrium spin systems with non-competing and competing external fields based on the TFD formulation. These results are summarized in Figs.~\ref{fig:01nc} $\sim$ \ref{fig:05}.
As is evident from the results, the values of $\hat{S}^{\alpha}_{\rm qe}$ are positive semi-definite. In particular, according to Eqs.~(12) and (13), the equilibrium-extended entropies, $\displaystyle \lim_{t \to \infty}\hat{S}^{\alpha}_{\rm qe}$, vanish at $H \to 0$. From these observation we infer that $\hat{S}^{\alpha}_{\rm qe}$ are well-defined measures of the quantum entanglement. This is also the condition that the entanglement measures of general quantum systems should satisfy.

Figures 1 and 3 show that the behaviors of $\hat{S}^{\rm nc}$, $\hat{S}_{\text{qe}}^{\rm nc}$, and  $b_{\text{qe}}^{\rm nc}$ differ little from those with no external fields \cite{Hashizume,Nakagawa}, and are not indicative of the Zeeman effect. This is due to the lack of competition between the interaction, $-J\bvec{S}_{\text{A}}\cdot \bvec{S}_{\text{B}}$, and the external fields, $-g\mu_BH\left(S_{\text{A}}^z+S_{\text{B}}^z\right)$, in the non-competing Hamiltonian, $\mathcal{H}^{\rm nc}$ in Eq.~\eqref{eq:1nc}. The spin inversion symmetry is still unbroken for the non-competing case. On the other hand, it is clear from Figs.~2 and 4 that the behaviors of $\hat{S}^{\rm c},~\hat{S}_{\text{qe}}^{\rm c}$, and  $b_{\text{qe}}^{\rm c}$ differ from those with no external field. Specifically, Figs.~2(d) and 4(d) show a ``twin-peaks'' oscillation with the same phase. This ``twin-peaks'' oscillation may be caused by the $H$ dependences of $\hat{S}_{\rm qe}^{\rm c}$ at $\epsilon =0$, which are shown in Fig.~\ref{fig:05}. These results show that the competing Hamiltonian, $\mathcal{H}^{\rm c}$ in Eq.~\eqref{eq:1a}, contains competing effects in a non-equilibrium system leading to the partial recovery of the spin inversion symmetry in the dissipative dynamics. This is recognized as the origin of the ``twin-peaks'' oscillation in the present model. 


In the present communication, the Hamiltonian was the most simple separation of Eq.\,\eqref{eq:last}. Using the replica trick and the Sherrington and Kirkpatrick mean-field solution \cite{Sherrington,Kirkpatrick,KS}, the free-energy and the order parameter of the equilibrium system of Eq.\,\eqref{eq:last} have been examined. However, these methods involve several serious problems, such as replica-symmetry breaking, ergodicity breaking, negative entropy at low temperatures, etc. On the other hand, the method presented here should be independent of the problem. Therefore, one can extend the method developed here to study the quantum entanglement in other spin systems. Furthermore, to clarify how general the appearance of ``twin-peaks '' oscillations is in spin systems, it is necessary to try the present method based on TFD in the above general spin-glass models. These extensions are currently under consideration.

\end{document}